\documentclass{article}

\usepackage{graphicx}
\usepackage{subfigure}
\usepackage{xcolor}
\usepackage{geometry}
\usepackage{caption}
\usepackage{amsmath}
\usepackage{comment}
\usepackage[colorlinks=true,linkcolor=black,citecolor=blue]{hyperref}
\usepackage{algorithm, algpseudocode}

\newtheorem{thm}{Theorem}[section]

\usepackage{multirow}
\title{A Separator-based Method for Generating Weakly Chordal Graphs}
\author{Md. Zamilur Rahman \\
	School of Computer Science \\
	University of Windsor \\
	Windsor, Canada \\
	\and Asish Mukhopadhyay \\
	School of Computer Science \\
	University of Windsor \\
	Windsor, Canada \\
	\and Yash P. Aneja \\
	Odette School of Business \\
	University of Windsor \\
	Windsor, Canada}
\date{}

\begin{document}
\maketitle{}

\begin{abstract}
We propose a scheme for generating a weakly chordal graph on $n$ vertices with $m$ edges. In this method, we first construct a tree and then generate an orthogonal layout (which is a weakly chordal graph on the $n$ vertices) based on this tree. In the next and final step, we insert additional edges to give us a weakly chordal graph on $m$ edges. Our algorithm ensures that the graph remains weakly chordal after each edge is inserted. The time complesity of an insertion query is $O(n^3)$ time and an  insertion takes constant time. On the other hand, a generation algorithm based on finding a 2-pair takes $O(nm)$ time using the algorithm of Arikati and Rangan~\cite{DBLP:journals/dam/ArikatiR91}.% or $O(n^{2.83})$ time using an algorithm due to Kratsch and Spinrad~\cite{DBLP:conf/soda/KratschS03}. 
\end{abstract}

\section{Introduction}\label{intro}
\begin{figure}[htb]
	\centering
	\subfigure[{\em $G$} \label{Fig-CGExample}]{\includegraphics[scale=.5]{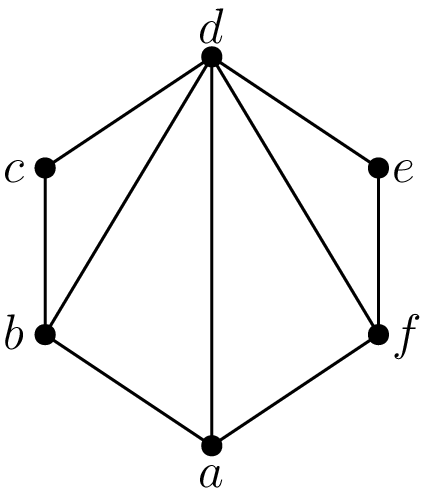}}\hspace{50pt}
	\subfigure[{\em $\overline{G}$} \label{Fig-CGCompExample}]{\includegraphics[scale=.5]{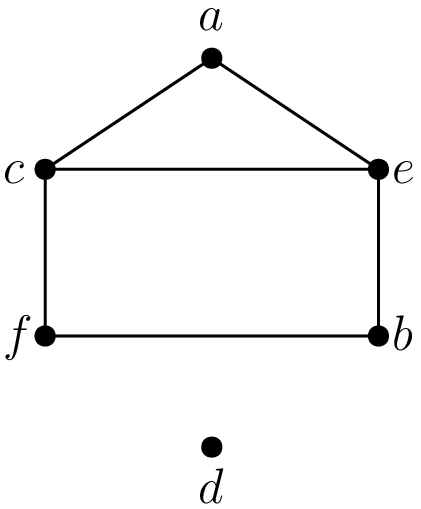}}
	\caption{{\em Complement of a chordal graph with a chordless 4-cycle}}
	\label{Fig-CGExamples}
\end{figure}

%Re-writing the Introduction
In this paper $G = (V, E)$ will be used to denote a graph with $|V| = n$ vertices and $|E| = m$ edges. $G$ is chordal if it has no chordless cycles of size four or more. However, as the following example shows (Figure~\ref{Fig-CGExamples}), its complement $\overline{G}$ can contain chordless 4-cycles.

%\begin{figure}[htbp]
%\centering
%\includegraphics[scale=.5]{images/addnlFigOne}
%\caption{{\em Complement of a chordal graph with a chordless 4-cycle}}
%\label{addnlFigOne}
%\end{figure}

\begin{figure}[htb]
	\centering
	\subfigure[{\em $G$} \label{Fig-WCGExample}]{\includegraphics[scale=.5]{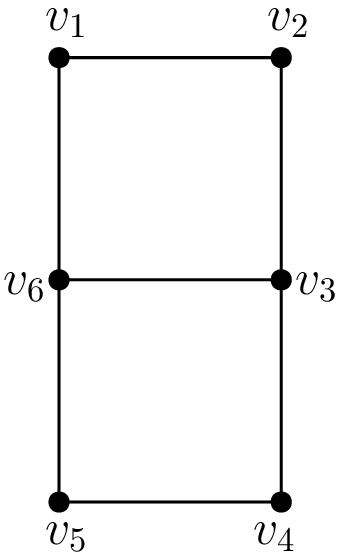}}\hspace{50pt}
	\subfigure[{\em $\overline{G}$} \label{Fig-WCGCompExample}]{\includegraphics[scale=.5]{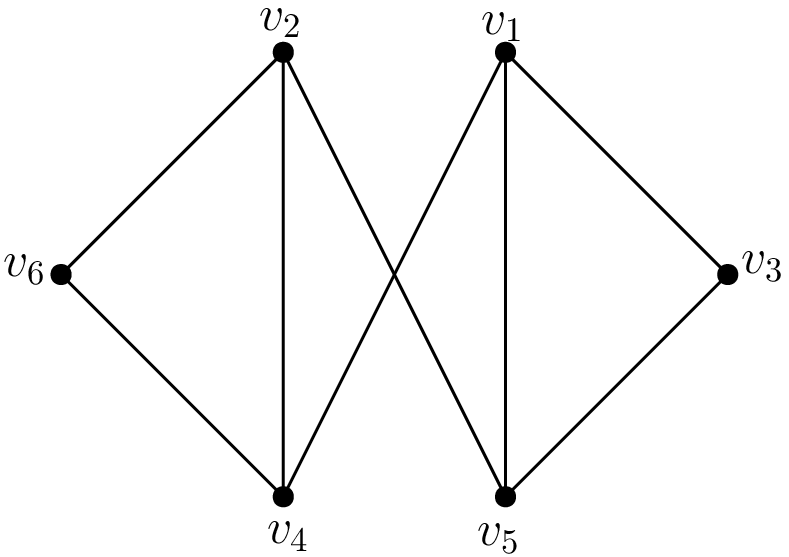}}
	\caption{{\em Weakly chordal graph}}
	\label{Fig-WCGExamples}
\end{figure}

This suggested the generalization of chordal graphs to weakly chordal (or weakly triangulated) graphs as those
graphs $G$ such that neither $G$ nor its complement $\overline{G}$ contains chordless
cycles of size five or greater \cite{DBLP:journals/jct/Hayward85}. 
From the symmetry of the definition it follows that $\overline{G}$ is also 
weakly chordal. Figure~\ref{Fig-WCGExamples} shows an example of a weakly chordal graph ($G$) along with its 
complement ($\overline{G}$). 

In this paper we address the algorithmic problem of generating weakly chordal graphs.
While the problem of generating graphs uniformly at random has received much 
attention (see~\cite{Tinhofer1990},~\cite{DBLP:journals/jal/Wormald84},~\cite{DBLP:journals/siamcomp/Wormald87},~\cite{DBLP:journals/jal/McKayW90}), little is known about this problem. There are many situations where we would like to generate instances of these to test algorithms for weakly chordal graphs. For instance, in~\cite{DBLP:journals/dmaa/MukhopadhyayRPG16} the authors generate all linear layouts of weakly chordal graphs. A generation mechanism can be used to provide input instances for this algorithm.

An edge $e$ in $G$ is {\em peripheral} if it is not the middle edge of a 
chordless path $P_4$ spanning four vertices of $G$.  In 
\cite{DBLP:journals/jgt/Hayward96}, Hayward proposed the following constructive 
characterization of weakly chordal graphs, based on the notion of a peripheral edge. 

\begin{thm}\label{thmwchordalgen}~\cite{DBLP:journals/jgt/Hayward96} 
	A graph is weakly chordal if and only if it can be generated in the following manner:
	\begin{enumerate}
		\item Start with an empty graph $G_0$.
		\item Repeatedly add an edge $e_j$ to the graph $G_{j-1}$ to create the graph $G_j$ such that $e_j$ is a peripheral edge of $G_j$.
	\end{enumerate}
\end{thm}

\noindent
This is analogous to a similar characterization for chordal graphs by Fulkerson 
and Gross~\cite{fulkerson1965}. No details were provided as to how to decide if 
an edge is peripheral and the complexity of the generation method. The proof of this 
theorem uses the notion of a two-pair and a generation method can be devised based on 
this. A pair of vertices $\{u, v\}$ in $G$ is a two-pair if the only chordless paths 
between the $u$ and $v$ are of length 2. Interestingly enough, a weakly chordal graph 
that is not a clique has a two-pair~\cite{DBLP:journals/gc/HaywardHM89}. Furthermore, let $\{u, v\}$ be a two-pair 
in an arbitrary graph $G$. Then $G + \{u, v\}$ is weakly chordal iff $G$ 
is weakly chordal~\cite{DBLP:journals/dam/SpinradS95}. We can then generate a weakly chordal graph on $n$ vertices 
by starting with a tree (which trivially satisfies the definition of weak chordality) 
and repeatedly find a two-pair $\{u, v\}$ and add to $G$ the edge joining $u$ to $v$. To
find a two-pair we can use an $O(mn)$ algorithm due to Arikati and Rangan~\cite{DBLP:journals/dam/ArikatiR91}. 
Unfortunately, this does not allow us to exploit the structural properties of a 
weakly chordal graph. 

Thus in this paper we propose a separator-based strategy that generalizes an algorithm due to Markenzon~\cite{DBLP:journals/anor/MarkenzonVA08} for generating chordal graphs and allows us to
exploit the structural properties of a weakly triangulated graph.   

The paper is organized thus. In the next section, we collect in one place some common graph-theoretic terminology used throughout the paper and also add a brief review of Markenzon's incremental method \cite{DBLP:journals/anor/MarkenzonVA08} for generating chordal graphs. The following section contains details of our algorithm, beginning with a brief overview. The final section contains some concluding remarks and suggestions for further research.

\subsection{Preliminaries}\label{preli}
%Graph Terminology
%Let $G=(V,E)$ be an undirected graph whose vertex set is $V$ and edge set $E$. 
For a given $G$, let $n = |V|$ and $m = |E|$. The neighborhood of a vertex $v$ is denoted as $N(v)$ where $N(v)=\{u\in V:(u,v)\in E\}$. The closed neighborhood of a vertex $v$, denoted by $N[v]$ is $N(v)\cup{v}$. The complement of a graph $G$, denoted by $\overline{G}$, contains the same set of vertices such that two vertices of $\overline{G}$ are adjacent iff they are not adjacent in $G$. The graph $\overline{G}$ in Fig~\ref{Fig-WCGCompExample} is the complement of the graph $G$ (Fig~\ref{Fig-WCGExample}).

For any vertex set $S\subseteq V$, the edge set $E(S)\subseteq E$ is defined to be $E(S) = \{(u,v)\in E | (u,v)\in S\}$. Let $G[S]$ denote the subgraph of $G$ induced by $S$, namely the subgraph $(S, E(S))$. In other words, an induced subgraph is a subset of the vertices of a graph $G$ together with any edges whose endpoints are both in this subset.

A path in a graph $G$ is a sequence of vertices $v_i, v_{i+1},\dots{.}, v_k$, where $\{v_j, v_{j+1}\}$ for $j=i, i + 1,\dots{.}, k-1$, is an edge of $G$. The first vertex is known as the start vertex, the last vertex is called the end vertex, and the remaining vertices in the path are known as internal vertices. A cycle is a closed path where the start vertex and the end vertex coincide. The size of a cycle is the number of edges in it. A clique in $G$ is a subset of vertices $(S\subseteq V)$ of $G$ such that the induced subgraph $G[S]$ is complete. A maximal clique is a clique that cannot be extended by including one more adjacent vertex. A graph on $n$ vertices that forms a clique on $n$ vertices is called a complete graph ($K_n$).

%Overview of Markenzon's method 
Based on the given $n$ and $m$, Markenzon's method starts by generating a labeled tree on $n$ vertices using Pr\"{u}fer's scheme~\cite{Prufer1918}. Next, an edge is inserted in each iteration to reach $m$. The algorithm picks a pair of vertices $\{u,v\}$ and checks whether $v$ is reachable from $u$ on the induced graph $G[S]$, where $S=V-I_{u,v}$ and $I_{u,v} = N(u) \cap N(v)$. If $G[S]$ is connected, then a path exists between $u$ and $v$ and thus $G+\{u,v\}$ is not chordal; otherwise, the augmented graph $G+\{u,v\}$ is chordal.
To make the search for a path efficient, the method reduces the set $S$ from $V-I_{u,v}$ to $N(x)-I_{u,v}$, for any $x\in I_{u,v}$. The correctness of the algorithm
was established by proving the following theorem. 

\begin{thm}~\cite{DBLP:journals/anor/MarkenzonVA08}\label{thmchordalInc}
	Let $G=(V,E)$ be a connected chordal graph and $u,v\in V,(u,v)\notin E$. The augmented graph $G+\{u, v\}$ is chordal if and only if $G[V-I_{u,v}]$ is not connected. %, where $I_{u,v} = N(u)\cap N(v)$.---$I_{u,v}$ already defined in the immediate above paragraph.
\end{thm}

\section{Generation of Weakly Chordal Graphs}\label{secgen}

\subsection{Overview of the Method}
%needs cleaning up
\begin{comment}
As the incremental method, we also take the number of vertices ($n$) and the number 
of edges ($m$) as input. Our proposed method starts with a tree and then generates an 
initial layout (which is a weakly chordal graph on $n$ vertices) from the tree. In 
the next and final step, the method adds the required number of edges to reach $m$ by 
maintaining the weak chordality property. The insertion of an edge into the weakly 
chordal graph ($G$) is symmetric to the deletion of an edge from the complement graph 
($\overline{G}$). Start with a complete graph and delete an edge in each iteration by 
following the same steps as for insertion.
\end{comment}

%In this section, we discuss the weakly chordal graph generation method. 
Just as in Markenzon's incremental method for chordal graphs, we take as input the number of vertices, $n$, and the number of edges, $m$, of the weakly chordal graph $G$ to be generated. Our proposed method starts with generating a tree. With the help of a tree, we construct an orthogonal layout made up of 4-cycles, each cycle corresponding to a node of this tree. We then insert additional edges into this initial layout to ensure that the final graph has $m$ edges. Trees are weakly chordal graphs and so is the initial orthogonal layout of the 4-cycles. Additional edges are introduced to reach the target value of $m$ by picking a pair of vertices $u$ and $v$ at random and checking if the augmented graph $G + \{u, v\}$  remains weakly chordal.  

%p_3 is not defined
If $u$ and $v$ have some common neighbors ($I_{u,v}\neq \emptyset$), we remove those common neighbors and check whether the removal of the common neighbors separates $u$ and $v$, leaving them in different components. We create an induced graph on a set of vertices called $AuxNodes$ that contain the neighbors of neighbors of $I_{u,v}$ to check whether $u$ and $v$ are in different components. If so we insert an edge between $u$ and $v$, else we search for shortest paths between $u$ and $v$. We do not insert the edge $\{u,v\}$ if the length of the shortest path is greater than $3$. Otherwise, we have to check for other conditions, such as single or multiple $P_3$ ($P_3$ denotes a path of length 3), forbidden configurations, alternate longer paths between $u$ and $v$ to decide whether the insertion of $\{u,v\}$ preserves weak chordality.

If there are no common neighbors ($I_{u,v}= \emptyset$) between $u$ and $v$, we apply the same rules to this case as when $I_{u,v}$ does not separate $u$ and $v$ (as explained in the previous paragraph). The only difference is that here we have to consider the entire graph to search for shortest paths between $u$ and $v$ but the other subsequent steps remain the same. All the phases of this generation method are explained in full details in the next subsections.

%what if the final tree has more than k nodes ?
\subsection{Phase 1: Generation of Tree}
%Note that each vertex of the tree contributes at least 2 new vertices to the final 
%layout. So our orthogonal layout has at least n vertices, even after introducing 
%a vertex to separate 2 adjacent degree 4 vertices.
%Let $k=\lceil{\frac{n}{2}}\rceil$. We generate a tree, $T$, on $k$ nodes that each node is of degree-4 at most as follows.
We generate a tree, $T$, on $\lceil{\frac{n}{2}}\rceil$ nodes that each node is of degree-4 at most as follows.
%Next we modify this tree 
%by introducing additionl  and two degree-4 nodes are not adjacent. 
Starting with a single node, we add new ones either by splitting an edge into two or joining a new node to one of the existing nodes, chosen at random. After $\lceil{\frac{n}{2}}\rceil$ nodes, have been added, we traverse the tree to check for adjacent degree-4 nodes. If such a pair is found, we separate them by introducing a new node adjacent to both. We define $k$ as the number of nodes in the tree ($T$). %That means we split the edge connecting the two nodes by inserting a new node. 

%how to delete nodes from initial layout? at first delete degree-1 node (if exist) and then delete degree-2 nodes. this make sure a connected component.
\subsection{Phase 2: Generation of Initial Layout}
\begin{figure}[htb]
	\centering
	\subfigure[\label{Fig-TreeLayoutExample-1}Tree ($T_1$)]{\includegraphics[scale=.67]{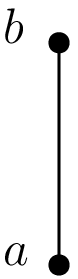}}\hspace{25pt}
	\subfigure[\label{Fig-TreeLayoutExample-2}Initial Layout ($G_1$)]{\includegraphics[scale=.67]{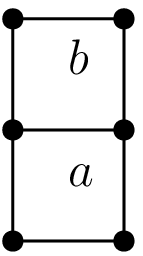}}\hspace{25pt}
	\subfigure[\label{Fig-TreeLayoutExample-3}Tree ($T_2$)]{\includegraphics[scale=.67]{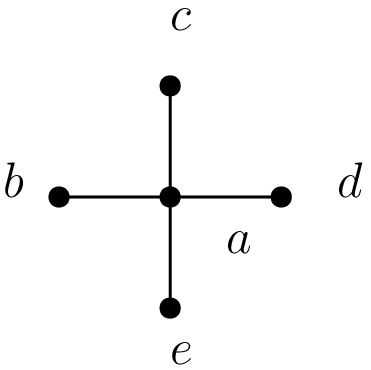}}\hspace{25pt}
	\subfigure[\label{Fig-TreeLayoutExample-4}Initial Layout ($G_2$)]{\includegraphics[scale=.67]{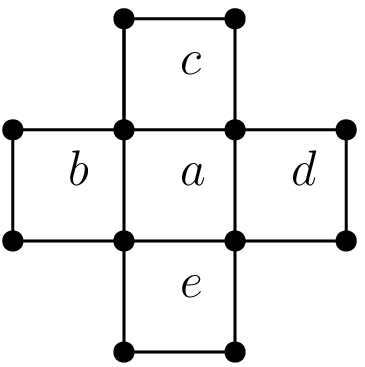}}\hspace{10pt}\\
	\subfigure[\label{Fig-TreeLayoutExample-5}Tree ($T_3$)]{\includegraphics[scale=.67]{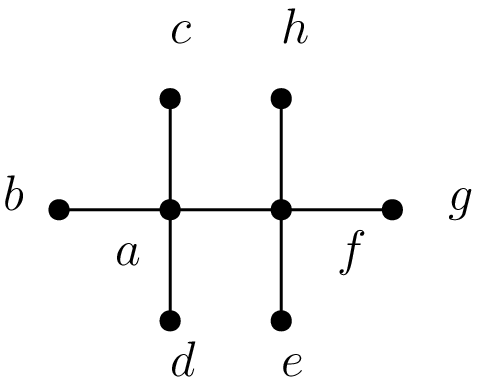}}\hspace{10pt}
	\subfigure[\label{Fig-TreeLayoutExample-6}Tree ($T_3^{\prime}$)]{\includegraphics[scale=.67]{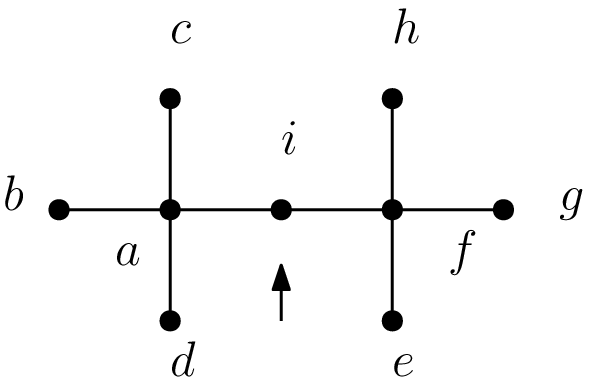}}\hspace{10pt}
	\subfigure[\label{Fig-TreeLayoutExample-7}Initial Layout ($G_3$)]{\includegraphics[scale=.67]{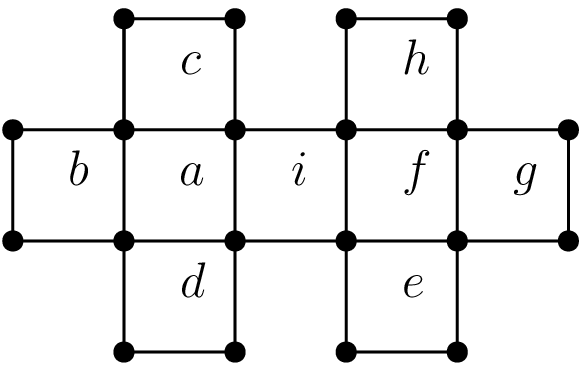}}
	\caption{{\em Tree to layout (four-cycles)}}
	\label{Fig-TreeLayoutExamples}%
\end{figure}

In this phase, we generate an orthogonal initial layout of $T$. For each node in $T$, we create a four-cycle. Figure~\ref{Fig-TreeLayoutExample-1} shows a tree $T$ with two nodes, $a$ and $b$. For node $a$ in the tree, we generate a four-cycle and for node $b$ we generate another four-cycle adjacent to the existing four-cycle because node $a$ is adjacent to node $b$ in the tree. The corresponding layout is shown in Figure~\ref{Fig-TreeLayoutExample-2}. We define this as the initial layout. For the example tree of Figure~\ref{Fig-TreeLayoutExample-3}, the corresponding initial layout is shown in  Figure~\ref{Fig-TreeLayoutExample-4}.

As explained in the first phase, after generating a tree, if two degree-$4$ nodes are adjacent, we insert a new node between them to separate them (see Figure~\ref{Fig-TreeLayoutExample-5},~\subref{Fig-TreeLayoutExample-6}, and~\subref{Fig-TreeLayoutExample-7}). Since a four-cycle has only four sides, two degree-$4$ nodes cannot be adjacent. As we are interested in generating proper weakly chordal graphs (and not just chordal graphs), we want to have four-cycles in the initial layout.  At this point, if the orthogonal layout has more than $n$ vertices, we delete vertices from four cycles to make it equal to $n$. To delete a vertex, we select vertices having degree at most two. If the number of edges in the initial layout is greater than or equal to the given $m$, the algorithm stops and returns the layout as a weakly chordal graph. Otherwise, it takes the difference of the number of edges and inserts that many edges in the initial layout to produce a weakly chordal graph. Now we denote the number of edges as $m^{\prime}$ in the initial orthogonal layout.
%NOTE: the generated intitial layout may have more than $m$ edges. By deleting a vertex from orthogonal layout we adjust $n$ but not $m$.

\subsection{Phase 3: Generation of Weakly Chordal Graph}
After generating an initial layout, we insert $(m-m^{\prime})$ edges to generate weakly chordal graphs. 
%(where $m$ is the input and $m^{\prime}$ is the number of edges in the initial layout). 
We divide the insertion of an edge $\{u,v\}$ into two cases. The first case considers when $I_{u,v}$ is non-empty and the second case is when $I_{u,v}$ is empty, where $I_{u,v} = N(u)\cap N(v)$, represents the set of common neighbors between $u$ and $v$. Let's discuss the case first when $I_{u,v}$ is non-empty.

\subsubsection{Case 1: $I_{u,v}$ is non-empty}\label{subsecMark}
Since $I_{u,v}$ is non-empty, we need to check whether the removal of $I_{u,v}$ separates $u$ and $v$. To do this, we create an induced graph, called $AuxGraph$ on the set of vertices $V-I_{u,v}$. Now we perform a breadth-first search on $AuxGraph$, starting at $u$ and if $v$ has not been reached, then the augmented graph $G+\{u,v\}$ is weakly chordal. Thus we can insert an edge $\{u,v\}$ in the weakly chordal graph $G$ because the removal of their common neighbors ($I_{u,v}$) in $AuxGraph$ leaves $u$ and $v$ in different components. Instead of searching for a path on this large set of vertices, we consider an induced graph on a smaller subset of vertices. In~\cite{DBLP:journals/anor/MarkenzonVA08}, the authors built the induced graph on the neighbors of any vertex $x$ in $I_{u,v}$ for the chordal graph. But here we need to consider the neighbors of neighbors of every vertex $x$ in $I_{u,v}$ to explore paths between $u$ and $v$. Formally the set of vertices included in $AuxNodes$ can be defined as $(\bigcup_{x\in I_{u,v}}(N(N(x\setminus\{u,v\}))\cup N(x)))\cup N(u)\cup N(v)-I_{u,v}$.%, $\forall x\in I_{u,v}$. 

\begin{figure}[htb]
	\centering
	\includegraphics[scale = .75]{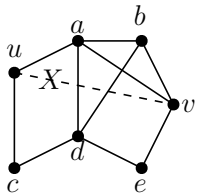}
	\caption{{\em Why neighbors of neighbors?}}
	\label{Fig-nOfNs}
\end{figure}

Consider the graph shown in Figure~\ref{Fig-nOfNs}. We want to introduce an edge between $u$ and $v$ which is not allowed because the insertion violates the property of the weakly chordal graph. There is only one common neighbor $I_{u,v}=\{a\}$ between $u$ and $v$. The $N(a)$ is $\{b,d\}$ and the induced graph on $\{u,v,b,d\}$ shows there is no path between $u$ and $v$ but there are longer paths ($u-c-d-e-v$ and $u-c-d-b-v$) exist between $u$ and $v$. These paths are discoverable if we take the neighbors of neighbors of $a$ and build $AuxGraph$ on $AuxNodes$ as defined above. The $N(N(a))$ includes $\{b,c,d,e\}$ and thus the $AuxNodes$ comprises $\{u,v,b,c,d,e\}$.

This is the first part of case 1 that we define as case 1.1. This idea is extended from~\cite{DBLP:journals/anor/MarkenzonVA08}, where the authors insert an edge $\{u, v\}$ in a chordal graph when $v$ has not been reached from $u$ during a breadth-first search on $AuxGraph$. What happens if $v$ is reachable from $u$ in a weakly chordal graph?

%In~\cite{DBLP:journals/anor/MarkenzonVA08}, the authors reject the insertion of an edge $\{u, v\}$ for chordal graph but 
In a weakly chordal graph we may be able to insert an edge $\{u, v\}$ even when a path exists between $u$ and $v$. In a chordal graph, the separators are cliques but in a weakly chordal graph, the separators may not necessarily cliques. Now we have to explore all the shortest paths of length 3 or more between $u$ and $v$. If there is a path longer than $P_3$ exists, we do not insert $\{u,v\}$ and choose another pair of vertices, otherwise, we check the other conditions for the possibility of insertion of $\{u, v\}$. We call this case 1.2. Again we divide this case 1.2 into two subcases: if there is a single $P_3$ then we called case 1.2.1 and otherwise case 1.2.2 for multiple $P_3$.

\begin{figure}[htb]
	\centering
	\subfigure[\label{Fig-NeighborsOfPath-1}]{\includegraphics[scale=.67]{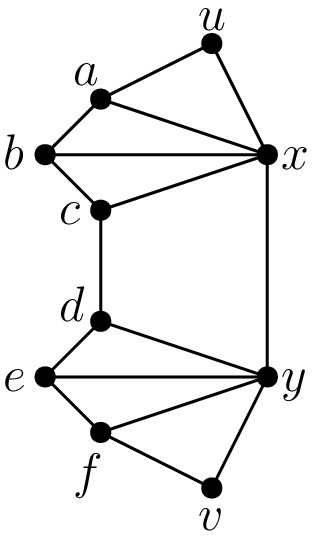}}\hspace{20pt}
	\subfigure[\label{Fig-NeighborsOfPath-2}]{\includegraphics[scale=.67]{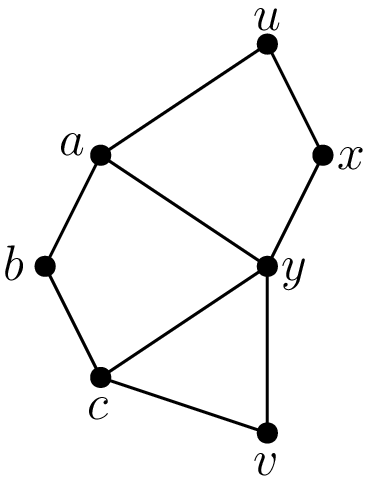}}
	\caption{{\em Neighbors of a path}}
%	\caption{{\em 
%		\subref{Fig-NeighborsOfPath-1} Each vertex on $p_i,i>3$ is a neighbor of either $x$ or $y$
%		\subref{Fig-NeighborsOfPath-2} There exists a vertex which is not a neighbor of $x$ or $y$
%	}}
	\label{Fig-NOfSP}%
\end{figure}

\textbf{Case 1.2.1:} We find the neighbors of the shortest path. If the neighborhood of the shortest path is empty, we insert an edge $\{u, v\}$, otherwise, we find the alternate paths in the entire graph. Instead of searching for alternate paths in the entire graph, we search for alternate paths on different induced graphs created on $\{N(N(SP\setminus\{u,v\}))\}$ (where $SP$ represents the shortest path between $u$ and $v$) by removing both the internal vertices of $SP$ or exactly one of them to find alternate paths between $u$ and $v$. Consider the two situations  shown in Figure~\ref{Fig-NOfSP} where $u-x-y-v$ is a path of length $3$. In Figure~\ref{Fig-NeighborsOfPath-1}, each vertex on $P_i$, $i>3$ is a neighbor of either $x$ or $y$ but in Figure~\ref{Fig-NeighborsOfPath-2} there exists a vertex ($b$) which is neither a neighbor of $x$ nor $y$. Thus we are interested in the neighbors of neighbors of the shortest path to discover other alternate paths.
The alternate paths may exist via some of the vertices in the shortest path and/or other vertices including in the neighbors or the neighbors of neighbors of the shortest path or disjoint path via the vertices not included in $SP$. If no alternate paths exist, we can insert an edge $\{u, v\}$, otherwise we do not insert $\{u,v\}$ and choose another pair of vertices. The algorithm~\ref{algoSSPWTG} explains the Case 1.2.1 (for a single shortest path). 

\begin{algorithm}[htb]
	\caption{SingleShortestPathSubCase}\label{algoSSPWTG}
	\begin{algorithmic}[1]
		\Require Single shortest path ($SP$, $u$, $v$)\Comment $\{SP={v_{0}-v_{1}-\dots-v_{l}}\}$
%		\Ensure Returns True if $G+\{u,v\}$ is weakly chordal
		\For {each $x$ in $SP$}
			\State Compute $N(x)$
		\EndFor
		\If { $N(SP)\setminus SP=\emptyset$}
			\State \textbf{Insert} edge $\{u,v\}$
		\Else
			\For {each $y$ in $N(SP\setminus\{u,v\})$}
				\State Compute $N(y)$ 
			\EndFor
			
			\State Create different induced graphs on $\{N(N(SP\setminus\{u,v\}))\}$ by removing both the internal vertices of $SP$ or exactly one of them %Note: different induced graphs are created by removing one or more inetrnal nodes to explore the other alternate paths. Not sure how to represent this step.
			\State Check for alternate longer path between $u$ and $v$
			\If {no alternate longer path exists}
				\State \textbf{Insert} edge $\{u,v\}$
			\EndIf
		\EndIf
	\end{algorithmic}
\end{algorithm}

\begin{figure}[htb]
	\centering
	\subfigure[\label{Fig-ForbiddenConfiguration-1}$G$]{\includegraphics[scale=.67]{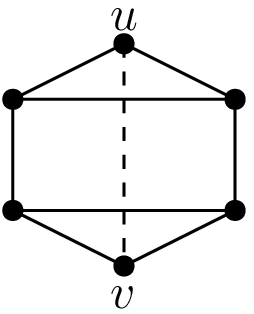}}\hspace{30pt}
	\subfigure[\label{Fig-ForbiddenConfiguration-2}$\overline{G}$]{\includegraphics[scale=.67]{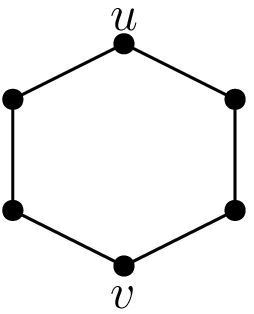}}
	\caption{{\em Forbidden configuration}}
	\label{Fig-ForbiddenConfigExamples}%
\end{figure}

\begin{figure}[htb]
	\centering
	\subfigure[\label{Fig-ForbiddenConfiguration-3}$G$]{\includegraphics[scale=.67]{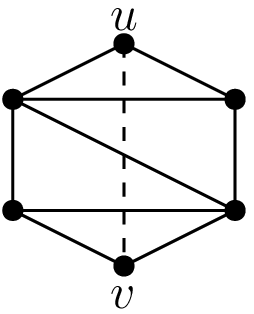}}\hspace{10pt}
	\subfigure[\label{Fig-ForbiddenConfiguration-4}$G$]{\includegraphics[scale=.67]{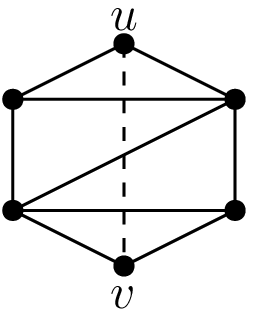}}\hspace{10pt}
	\subfigure[\label{Fig-ForbiddenConfiguration-5}$G$]{\includegraphics[scale=.67]{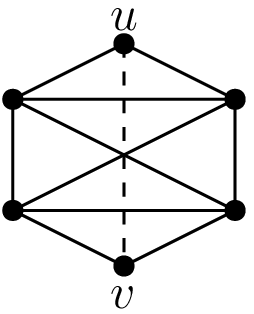}}\hspace{10pt}
	\subfigure[\label{Fig-ForbiddenConfiguration-6}$G$]{\includegraphics[scale=.67]{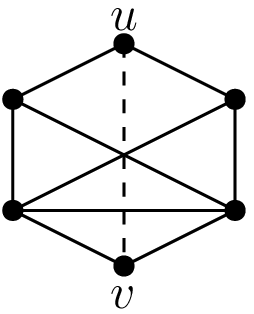}}\hspace{10pt}
	\subfigure[\label{Fig-ForbiddenConfiguration-7}$G$]{\includegraphics[scale=.67]{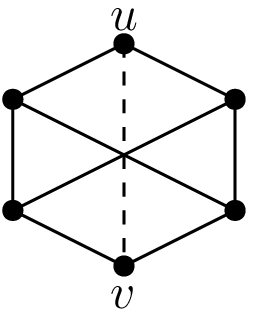}}\hspace{10pt}
	\subfigure[\label{Fig-ForbiddenConfiguration-8}$G$]{\includegraphics[scale=.67]{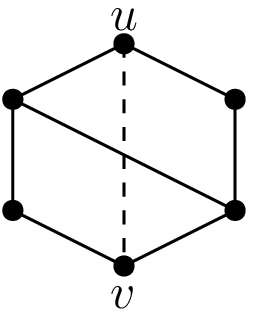}}\hspace{10pt}
	\subfigure[\label{Fig-ForbiddenConfiguration-9}$G$]{\includegraphics[scale=.67]{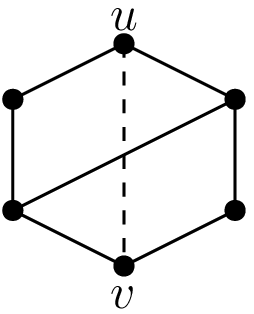}}
	\caption{{\em Other allowed configurations}}
	\label{Fig-OtherConfigExamples}%
\end{figure}

\textbf{Case 1.2.2:} In the case of multiple shortest paths, first we need to check for forbidden configurations. To define a forbidden configuration, let's consider the graphs shown in Figure~\ref{Fig-ForbiddenConfigExamples}. In Figure~\ref{Fig-ForbiddenConfiguration-1}, we can insert an edge $\{u,v\}$ because it does not create any chordless five cycles or more in the graph $G$, but it creates a chordless six cycle in the complement graph as illustrated in Figure~\ref{Fig-ForbiddenConfiguration-2}. As we know from the definition of the weakly chordal graph, a graph $G$ is weakly chordal if neither $G$ nor its complement $\overline{G}$ contains a chordless cycle of size $5$ or more. But $G+\{u,v\}$ creates a chordless six cycle in $\overline{G}$. Thus, we cannot add $\{u,v\}$ in $G$. The forbidden configuration checking is required when two or more shortest $P_3$ paths exist between $u$ and $v$ and the checking has to be done by considering all pairwise disjoint shortest paths. Figure~\ref{Fig-OtherConfigExamples} illustrates other configurations where adding $\{u,v\}$ to $G$ may be allowed.

\begin{figure}[htb]
	\centering
	\subfigure[\label{Fig-PathDisjoint}Disjoint path]{\includegraphics[scale=.65]{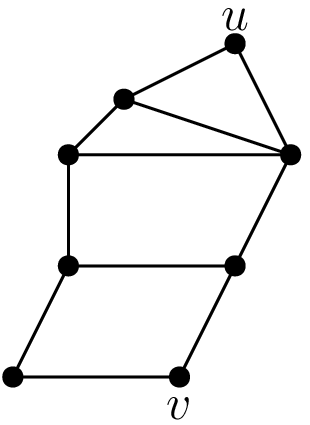}}\hspace{20pt}
	\subfigure[\label{Fig-PathShared}Shared path]{\includegraphics[scale=.65]{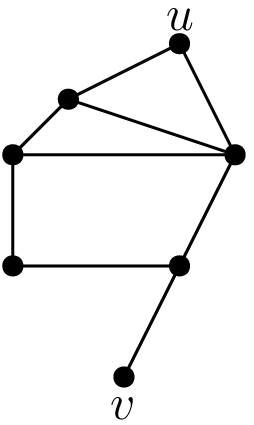}}
	\caption{{\em Different paths between $u$ and $v$}}
	\label{Fig-PathExamples}%
\end{figure}

After checking forbidden configurations, the next step is to check for alternate longer paths between $u$ and $v$. Like case 1.2.1, instead of searching for alternate longer paths in the entire graph we do the search on the different induced graphs created on $\{N(N(allSP\setminus\{u,v\}))\}$ (where $allSP$ represents all the shortest paths between $u$ and $v$ containing the set of vertices in the paths including two endpoints) by removing all the internal vertices of $allSP$ or exactly one of them to find alternate paths between $u$ and $v$. The alternate paths may exist via some of the vertices in the shortest path and/or the other vertices including in the neighbors/the neighbors of neighbors of the shortest path or disjoint path via the vertices not included in $allSP$. These two scenarios are illustrated in Figure~\ref{Fig-PathExamples}.

\begin{algorithm}[H]
	\caption{MultipleShortestPathSubCase}\label{algoMSPWTG}
	\begin{algorithmic}[1]
		\Require Multiple shortest paths ($allSP=\{SP_{0}, SP_{1},\dots, SP_{j}\}$, $u$, $v$)\Comment $\{SP_{i}={v_{0}-v_{1}-\dots-v_{l}}\}$
		\If {no forbidden configuration}
			\For {each SP}
				\For {each $x$ in $SP$}
					\State Compute $N(x)$
				\EndFor
			\EndFor
			\For {each $y$ in $N(allSP\setminus \{u,v\})$}
				\State Compute $N(y)$ 
			\EndFor
			\State Create different induced graphs on $\{N(N(allSP\setminus\{u,v\}))\}$ by removing both the internal vertices of $SP$ or exactly one of them %Note: different induced graphs are created by removing one or more inetrnal nodes to explore the other alternate paths. Not sure how to represent this step.
			\State Check for alternate longer path between $u$ and $v$
			\If {no alternate longer path exists}
				\State \textbf{Insert} edge $\{u,v\}$
			\EndIf
		\EndIf
	\end{algorithmic}
\end{algorithm}

\begin{algorithm}[H]
	\caption{FromInitLayoutToWCG}\label{algoMainWTG}
	\begin{algorithmic}[1]
		\Require An initial layout $G=(V,E)$
		\Ensure A weakly chordal graph $G+\{u,v\}$
		\State $p=0$; $q=m-m^{\prime}$\Comment{$m^{\prime}$ is the no. of edges in the initial layout}
		\While {$p < q$ }
			\State Choose a pair of vertices $u$ and $v$
			\If {$\{u,v\}$ exists}
				\State \textbf{continue}
			\Else
				\State Compute $N(u)$ and $N(v)$ \label{nunv}
				\State Compute $I_{u,v} = N(u)\cap N(v)$ \label{iuv}
				\If {$I_{u,v}$ is non-empty} \Comment  case 1: $I_{u,v}\neq\emptyset$
					\For {each $x$ in $I_{u,v}$}
						\State Compute $N(x)$
					\EndFor
					\For {each $y$ in $N\big(x\setminus \{u,v\}\big)$}
						\State Compute $N(y)$
					\EndFor
					\State Create $AuxGraph$ on $(\bigcup_{x\in I_{u,v}}(N(N(x\setminus\{u,v\}))\cup N(x)))\cup N(u)\cup N(v)-I_{u,v}$
					\State Find whether there is a path exist between $u$ and $v$ in $AuxGraph$
					\If {$v$ is not reachable from $u$}\Comment case 1.1
						\State \textbf{Insert} edge $\{u,v\}$
					\ElsIf {Find all the shortest paths between $u$ and $v$}\Comment case 1.2
						\If {all shortest paths are of length 3 ($P_3$)}
							\If {there is a single shortest path between $u$ and $v$}\Comment case 1.2.1
								\State {\textbf{call singleShortestPathSubCase}}
							\ElsIf {there are multiple shortest paths between $u$ and $v$}\Comment case 1.2.2
								\State {\textbf{call multipleShortestPathsSubCase}}
							\EndIf
						\EndIf
					\EndIf
				\Else \Comment  case 2: $I_{u,v}=\emptyset$
					\If {all shortest paths are of length 3 ($P_3$)}
						\If {there is a single shortest path between $u$ and $v$}\Comment case 2.1
							\State {\textbf{call singleShortestPathSubCase}}
						\ElsIf {there are multiple shortest paths between $u$ and $v$}\Comment case 2.2
							\State {\textbf{call multipleShortestPathsSubCase}}
						\EndIf
					\EndIf
				\EndIf
			\EndIf
		\EndWhile
	\end{algorithmic}
\end{algorithm}

\subsubsection{Case 2: $I_{u,v}$ is empty}
In this case, there are no common neighbors of $u$ and $v$ but a path exists between $u$ and $v$. This case can be solved in a similar way as for case 1.2. Here the $AuxGraph$ is the entire graph because $I_{u,v}$ is empty. If any of the paths is greater than $P_3$, we do not insert $\{u,v\}$ and choose another pair of vertices. Otherwise, we apply the same single shortest path case 1.2.1 and multiple shortest path case 1.2.2 to check if the addition of $\{u,v\}$ keeps the graph weakly chordal or not. Here we define the single shortest path case as case 2.1 and multiple shortest paths case as case 2.2. The details of all the above cases are formally described in algorithm~\ref{algoMainWTG}.

\subsubsection{An Example}
\begin{figure}[htb]
	\centering
	\subfigure[\label{Fig-WCGInitialExample-1}Tree ($T_1$)]{\includegraphics[scale=.75]{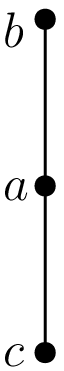}}\hspace{30pt}
	\subfigure[\label{Fig-WCGInitialExample-2}Initial Layout ($G_1$)]{\includegraphics[scale=.75]{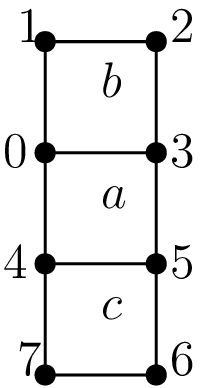}}\hspace{30pt}
	\subfigure[\label{Fig-WCGInitialExample-3}WCG($G_1^{\prime}$)]{\includegraphics[scale=.75]{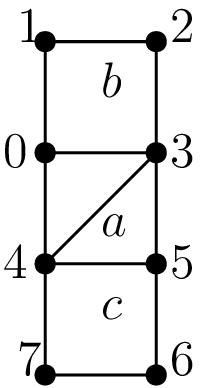}}\hspace{30pt}
	\subfigure[\label{Fig-WCGInitialExample-4}WCG($G_1^{\prime\prime}$)]{\includegraphics[scale=.75]{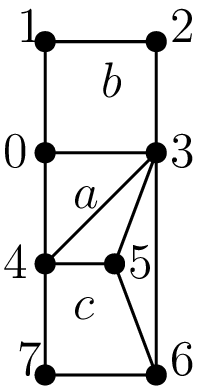}}
	\caption{{\em Tree to weakly chordal graph}}
	\label{Fig-WCGInitialExamples}%
\end{figure}

Figure~\ref{Fig-WCGInitialExamples} shows an example for the generation of weakly chordal graph for $n=8$ and $m=12$. All the phases are shown in Figure~\ref{Fig-WCGInitialExamples}. Figure~\ref{Fig-WCGInitialExample-1} shows the tree on $\lceil{\frac{n}{2}}\rceil$ nodes. There are three nodes in the tree and for each node in the tree, we create an orthogonal initial layout which is shown in Figure~\ref{Fig-WCGInitialExample-2} where $n=8$ and $m^{\prime}=10$. We assign the labels of vertices as $0\dots 7$ in the initial layout. We need to insert $(m-m^{\prime})=2$ more edges in the initial layout to generate a weakly chordal graph for the given input. 

Assume we want to insert edge $\{3,4\}$. The insertion of an edge between $3$ and $4$ falls into the case where $I_{3,4}$ is non-empty. Since the removal of $I_{3,4}$ leaves vertices $3$ and $4$ in two different components, so we can insert an edge between vertices $3$ and $4$ according to case 1.1 as shown in Figure~\ref{Fig-WCGInitialExample-3}. Now we want to insert edge $\{3,6\}$. The insertion of an edge between $3$ and $6$ falls into the case 1.2.1 because the removal of their common neighbor $\{5\}$ does not separate $3$ and $6$. The $N(5)$ is $\{3,4,6\}$ and the $N(N(\{3,4,6\})\setminus\{3,6\})$ is $\{0,5,7,3\}$. Now we create the $AuxGraph$ on the $AuxNodes=\{0,2,3,4,6,7\}$ and then we search for the paths from $3$ to $6$. There is a single shortest path $SP=\{3,4,7,6\}$ exist that means it falls into case 1.2.1. We compute $N(SP)=\{0,2,4,3,6,7\}$ and the $N(SP)\setminus SP\neq \emptyset$, so we need to compute $\{N(N(SP\setminus\{3,6\}))\}$ which is $\{3,4,0,7,6\}$ and create induced graph on these set of vertices. To discover disjoint or shared alternate paths from $3$ to $6$ we create different induced graphs by removing some vertices from the above-computed set of vertices. But there is no such path exists and we can insert an edge between $3$ and $6$.

%Why 3k + 1 ? Each new 4-cycle introduces 3 new edges. The first one itroduces 4. 
%Therefore 4 + 3(k-1) = 3k + 1
\subsection{Time Complexity}
The tree generation is the first phase in the proposed approach and can be constructed in time $O(k)$ where $k$ is the number of nodes in a tree. For $k$ nodes in the tree, we insert $3k+1$ edges in the layout and each edge insertion can be done in constant time. So the initial layout can be generated in $O(k)$ time.

In the third phase, the pair of vertices $\{u,v\}$ is chosen at random to insert an edge between them. Two types of failures may arise. One is that the pair of vertices $\{u,v\}$ corresponds to an existent edge and the other is the addition of $\{u,v\}$ violates weak chordality property. To avoid the first type of failures either we can maintain a list of edges (those are present in the complement graph) not present in the current graph or we can check the existence of an edge in constant time by maintaining the adjacency matrix representation of the current graph.

%step 8 need to be changed.
Computing the neighborhood of a vertex takes linear time. Thus step 7 takes linear time. By amortizing over the edges we can see that step 8 takes linear time in the number of edges $m$ and the number of vertices $n$. Thus each of the neighborhood computations take linear time. The $AuxGraph$ can be created in linear time on the number of vertices in the $AuxNodes$. Using breadth-first search (BFS), we find whether there is a path exist between $u$ and $v$ in linear time. To find all $P_3$ shortest paths we also use BFS by keeping track of all predecessors between $u$ and $v$. All the operations including neighborhood computations, induced graph creations, and finding alternate longer paths take linear time (see step 23 and step 32). 

For steps 25 and 34, the forbidden configuration checking is required when two or more shortest $P_3$ paths exist between $u$ and $v$ and the checking has to be done by considering all pairwise disjoint shortest paths. The checking is performed by the presence of one of the cross edges (an edge from one of the internal nodes from one shortest path to the alternate internal node of another shortest path) on the induced graph on a pair of disjoint shortest paths. There are at most $O(n^3)$ pairs of mutually disjoint $P_3$ paths between $u$ and $v$ and thus the step takes $O(n^3)$ time (see Appendix). 
%To discover disjoint or shared paths we create different induced graphs on the neighborhood and/or the neighborhood of neighborhood of the shortest paths. 
We use BFS to find an alternate longer path between $u$ and $v$. Since there are at most $O(n^2)$ multiple $P_3$ paths between $u$ and $v$ and thus in total we spend $O(n^2)$ time. In the worst case, an insertion query runs in $O(n^3)$ time.

The insertion of $\{u,v\}$ is performed in time $O(1)$ because we need to insert $u$ in the neighborhood of $v$ and $v$ in the neighborhood of $u$.

\section{Conclusions}\label{secconclusions}
We have proposed a method for the generation of weakly chordal graphs. We have implemented the proposed algorithm in Python. An interesting open problem is to investigate how to generate weakly chordal graphs uniformly at random. This requires coming up with a scheme for counting the number of labeled weakly chordal graphs on $n$ vertices having $m$ edges. If we could also determine the probability distribution underlying our algorithm for generating weakly chordal graphs, then we could compute the relative entropy between the two distributions to estimate how close our algorithm is to generating weakly chordal graphs uniformly at random.
%Another interesting problem is to find a characterizations of  various  to analyze the varieties of the weakly chordal graphs.

\section{Appendix}\label{app}
\begin{figure}[htbp]
\centering
\includegraphics[scale=.6]{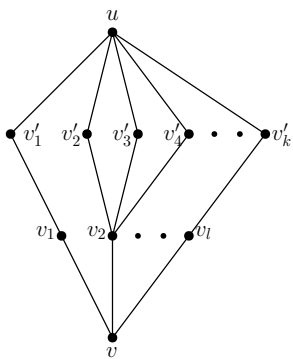}
\caption{{\em $P_3$ paths between $u$ and $v$}}
\label{Fig-PathComputation}
\end{figure}

We are interested in finding all $P_3$ shortest paths between $u$ and $v$ (see Figure~\ref{Fig-PathComputation}). Suppose there are $n$ vertices in the graph. Let $\{v_1,v_2,\dots,v_l\}$ be the set of vertices adjacent to $v$ that lie on the $P_3$ shortest paths between $u$ and $v$. Likewise, let $\{v_1^{\prime},v_2^{\prime},v_3^{\prime},v_4^{\prime}\dots v_k^{\prime}\}$ be the set of vertices adjacent to $u$ that lie on these shortest paths. Thus $lk$ is the upper bound of the number of paths between $u$ and $v$. Since $(l+k)\leq (n-2)$, then $lk=O(n^2)$.

%The $lk$ become maximum when $l$ and $k$ are equal and the value of $l$ and $k$ can be computed as $\frac{n-2}{2}$ and thus $lk = (\frac{n-2}{2})(\frac{n-2}{2})$. So there are $n^2$ number of $P_3$ shortest paths between $u$ and $v$. 

Let $d_i$ be the degree of $v_i$ relative to the vertices $\{v_1^{\prime},v_2^{\prime},v_3^{\prime},v_4^{\prime}\dots v_k^{\prime}\}$, for $i=1,\dots,l$. Then the number of pairs of edge-disjoint paths between $u$ and $v$ is given by $PathCount=\sum_{i\neq j}d_id_j$. Let $\sum d_i=t$. Now it follows from the equality $2l\sum d_id_j=(l-1)(\sum d_i)^2-\sum_{i\neq j} (d_i-d_j)^2$ that $PathCount$ is maximum when all $d_i$'s are equal. Therefore an upperbound on $PathCount$ is $l^2.\frac{t}{l}=lt$. Since $t=O(lk)$, we have $PathCount=O(l^2k)$.

%($lk$ represents the sum of all indegrees (we use indegree to refer all the outgoing edges from $k$ vertices to $l$ vertices) distributed among $l$ vertices).

%Thus $PathCount$ is $l^2.\frac{lt}{l}=l^2t$. Hence there are $n^3$ number of pairwise disjoint paths between $u$ and $v$.

%from The term $\sum_{i\neq j} d_id_j$ given $\sum d_i$ is maximum when $2l\sum d_id_j=(l-1)(\sum d_i)^2-\sum_{i\neq j} (d_i-d_j)^2$ holds. Because there are at most $l^2$ number of edges incident on $k$ vertices and $lt$ represents the sum of $t$ indegrees (we use indegree to refer all the outgoing edges from $k$ vertices to $l$ vertices) distributed among $l$ vertices.  Hence there are $n^3$ number of pairwise disjoint paths between $u$ and $v$.
%$\frac{lt}{l}=l^2t$. 
%$2l\sum d_id_j=(l-1)(\sum d_i)^2-\sum_{i\neq j} (d_i-d_j)^2$

%\bibliography{weaklyCGRefs}
%\bibliographystyle{plain}

%reference added in tex because of the requirement to arxiv

\end{document}